# DEVELOPMENT OF QUANTUM CIRCUITS FOR PERCEPTRON NEURAL NETWORK TRAINING, BASED ON THE PRINCIPLES OF GROVER'S ALGORITHM

Cesar Borisovich Pronin

Andrey Vladimirovich Ostroukh

MOSCOW AUTOMOBILE AND ROAD CONSTRUCTION STATE TECHNICAL UNIVERSITY (MADI)., 64, Leningradsky prospect, Moscow, Russia

**Abstract:** This paper highlights a practical research of the possibility of forming quantum circuits for training neural networks. The demonstrated quantum circuits were based on the principles of Grover's Search Algorithm. The perceptron was chosen as the architecture for the example neural network. The multilayer perceptron is a popular neural network architecture due to its scalability and applicability for solving a wide range of problems.

## Introduction

To demonstrate the concept of using Grover's algorithm for training neural networks we shall use an example neural network (Fig. 1), which, scaled up, could be a part or a prototype of a more complex neural network, designed, for example, to compare objects on different images. The given example of a neural network consists of two input neurons, one hidden and one output. The letters $I$ denote the input neurons, the letter $H$ - the hidden neuron, and the letter $w$ - the weights of the synapses (connections between neurons).

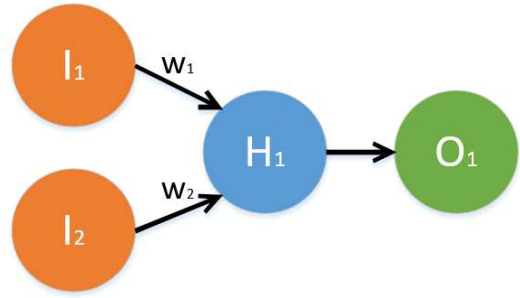

Fig. 1. Perceptron example

The input data of the hidden neuron is formed as the sum of input neuron values multiplied by corresponding synapse weights (1).

$$H_{1\ input} = I_1 w_1 + I_2 w_2 \qquad (1)$$

The input data is transferred to the activation function (2) which forms the output data of the hidden neuron. A similar operation is repeated for the entire network, until the output neuron is reached. The initial values of the weights $w_1$ and $w_2$ are usually set randomly. The choice of using a certain activation function often depends on the purpose of the neural network.

$$H_{1\ output} = f_{activation}(H_{1\ input}) \qquad (2)$$

Since the task of comparing objects was chosen as an application example, the following neuron activation function could be proposed:

```
If  H_{1 input} ≥ Ac ,  then H_{1 output} = 1;

Else H_{1 output} = 0;
```

The process of constructing a neural network also includes the process of its training. Training consists of adjusting the synapse weights of the network to obtain required output values [1]. The training process of our neural network can be represented as a solution to inequality (3), in which the value of $Ac$ is known and is the threshold value required to obtain "1" as the output of the hidden neuron.

$$I_1 w_1 + I_2 w_2 \geq Ac \qquad (3)$$

To simplify the quantum circuit, we will model the situation where values of $I_1$ and $I_2$ are strictly equal to either 0 or 1 (the case when the original pixels of the image can only be black or white) and find only the minimum values of weights that satisfy the condition in the activation function (4).

$$I_1 w_1 + I_2 w_2 = Ac \qquad (4)$$

Let's solve equation (4) as an equation with two unknown variables $w_1$ and $w_2$, by using Grover's search algorithm. The output of the circuit on Fig. 3 would consist of suitable values for $w_1$ and $w_2$, for inputs $I_1$ and $I_2$. These datasets are shown in Table 1.

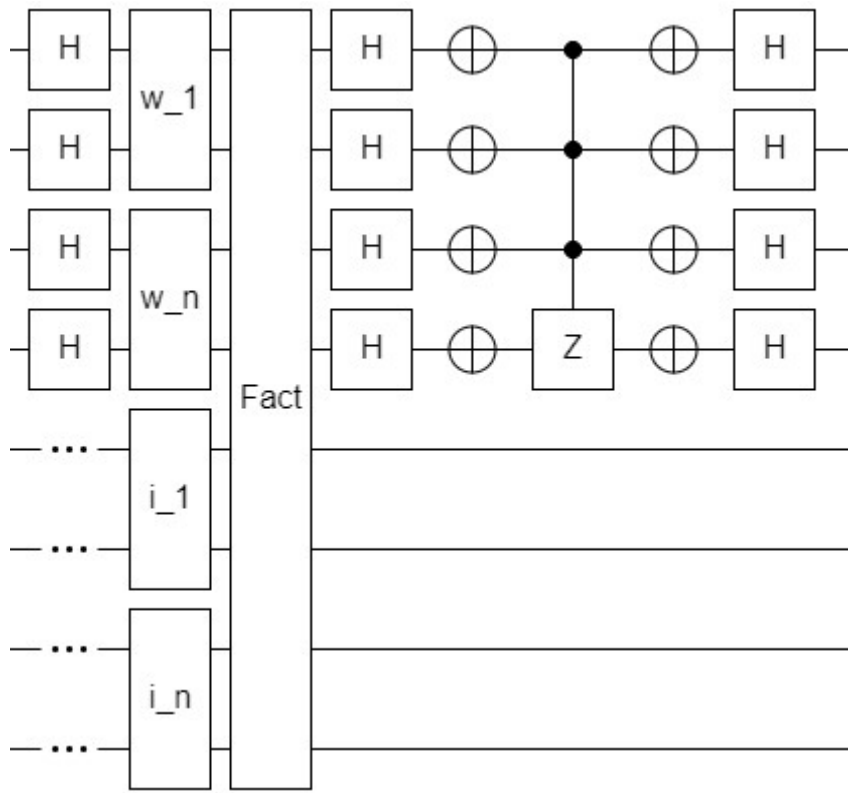

*Fig. 2. The proposed concept for constructing perceptron training circuits, using the Grover quantum algorithm*

The proposed quantum perceptron training circuit (Fig. 2) is based on a Grover's algorithm, in which the structure of the oracle function (function that serves as a search criteria of the Grover's algorithm) is built based off the activation functions of hidden neurons ($F_{act}$ – Fig. 2) and the structure of the neural network itself. The main objective in constructing this training function, is to provide a condition, which could be used in this oracle function to select suitable weight values. The input data of this quantum circuit will consist of input neuron values $I_1..I_n$. The output data of the algorithm will consist of all possible combinations of weight values $w_1..w_n$ that satisfy the activation function.

Because Grover's algorithm is used to train the perceptron, based on the description of the algorithm [2-4], the training result as a set of weights $w_1..w_n$ can be obtained by repeating $N_G$ Grover iterations (the oracle and amplitude amplification functions), where *n* is the number of quantum bits, used by the algorithm (occupied by the amplitude amplification function), and *l* is the number of expected solutions. In our case, *l* is the number of weight sets that satisfy the activation function (if known). If the number *l* is unknown, it could be equaled to 1, and then the number of Grover iterations should be decreased until the maximum accuracy of the results is achieved.

$$N_G = \left\lceil \frac{\pi}{4} * \sqrt{\frac{2^n}{l}} \right\rceil \tag{5}$$

The classical method for training multilayer perceptron networks is backpropagation. It focuses on calculating the error value and changing the synapse weights sequentially in direction from the output to the input neurons, in order to reduce this error value.

Comparing the performance of different training methods is a rather difficult task, due to complexity in predicting the number of iterations required by a classical algorithm to find the minimum error. In the paper "Quadratic Quantum Speedup for Perceptron Training" [5] that highlights the possibility of using another method based on Grover's algorithm for use in machine learning, the authors point out the possibility of achieving quadratic acceleration in perceptron training.

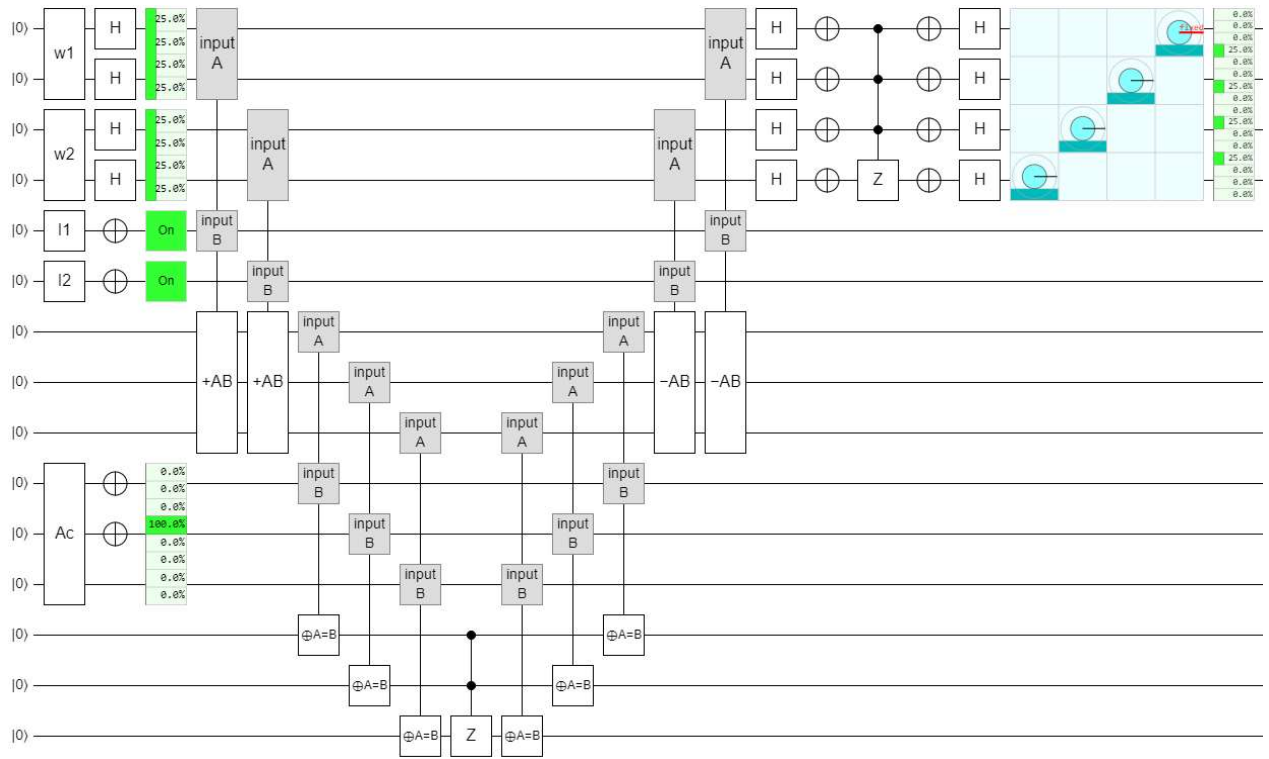

*Fig. 3. Perceptron training quantum circuit for $Ac = 011_2 = 3_{10}$ that utilizes the principles of Grover's algorithm, based on the concept on fig. 2*

Table 1. Output values *w* given for different inputs *I, Ac* by the circuit on fig. 3 (binary numbers)

| $w_1$ | $w_2$ | $I_1$ | $I_2$ | Ac |
|---|---|---|---|---|
| 00 | 00 | 0 | 0 | |
| 11 | 00 | 1 | 0 | |
| 11 | 01 | | | |
| 11 | 10 | | | |
| 11 | 11 | | | |
| 00 | 11 | 0 | 1 | 011 |
| 01 | 11 | | | |
| 10 | 11 | | | |
| 11 | 11 | | | |
| 11 | 00 | 1 | 1 | |
| 01 | 10 | | | |
| 10 | 01 | | | |
| 00 | 11 | | | |
| | | | | |

If the *Ac* threshold gets set to a higher value, the number of output weight combinations decreases.

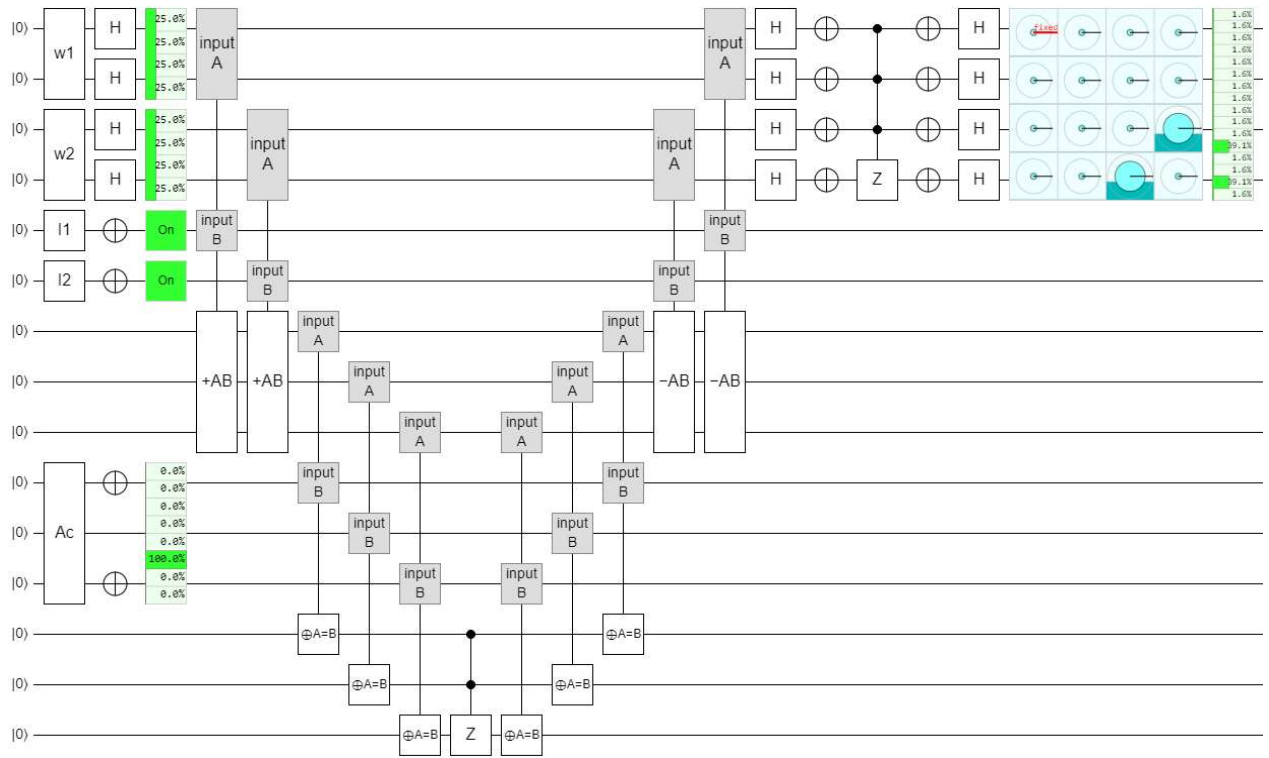

*Fig. 4. Perceptron training quantum circuit for $Ac = 101_2 = 5_{10}$ that utilizes the principles of Grover's algorithm*

Table 2. Output values *w* given for different inputs *I, Ac* by the circuit on Fig. 4 (binary numbers)

| $w_1$ | $w_2$ | $I_1$ | $I_2$ | Ac |
|---|---|---|---|---|
| 00 | 00 | 0 | 0 | 101 |
| 00 | 00 | 1 | 0 | |
| 00 | 00 | 0 | 1 | |
| 10 | 11 | 1 | 1 | |
| 11 | 10 | | | |

When performing circuits on Fig. 2-4, it is necessary to execute the algorithm, gradually changing the values of the input neurons $I$ in order to determine appropriate weight sets. This may not be the most efficient solution in terms of using Grover's algorithm, so we will consider an alternative circuit.

Let's suppose that the inequality $I_1 w_1 + I_2 w_2 \geq Ac$ has four unknown variables: $I_1, I_2, w_1, w_2$. To solve it, we will slightly change the concept of the circuit, from Fig. 2 to the concept on Fig. 5. Then the output of our example circuit should form all sets of values $I_1, I_2, w_1, w_2$ suitable for training the neural network, selected by checking all possible input states $I_1, I_2, w_1, w_2$ (table 1).

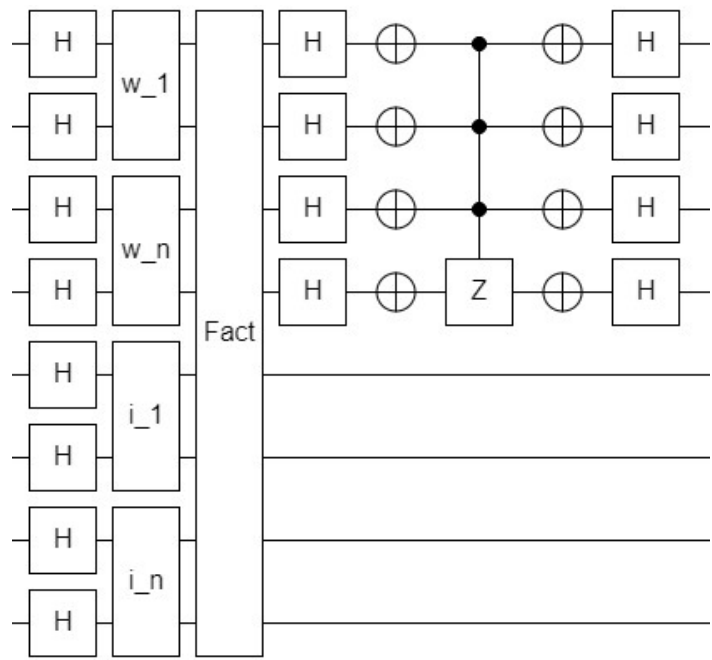

*Fig. 5. The proposed concept of constructing perceptron training circuits. The circuit analyzes all possible sets of input states $I_1, I_2, w_1, w_2$ and selects suitable combinations*

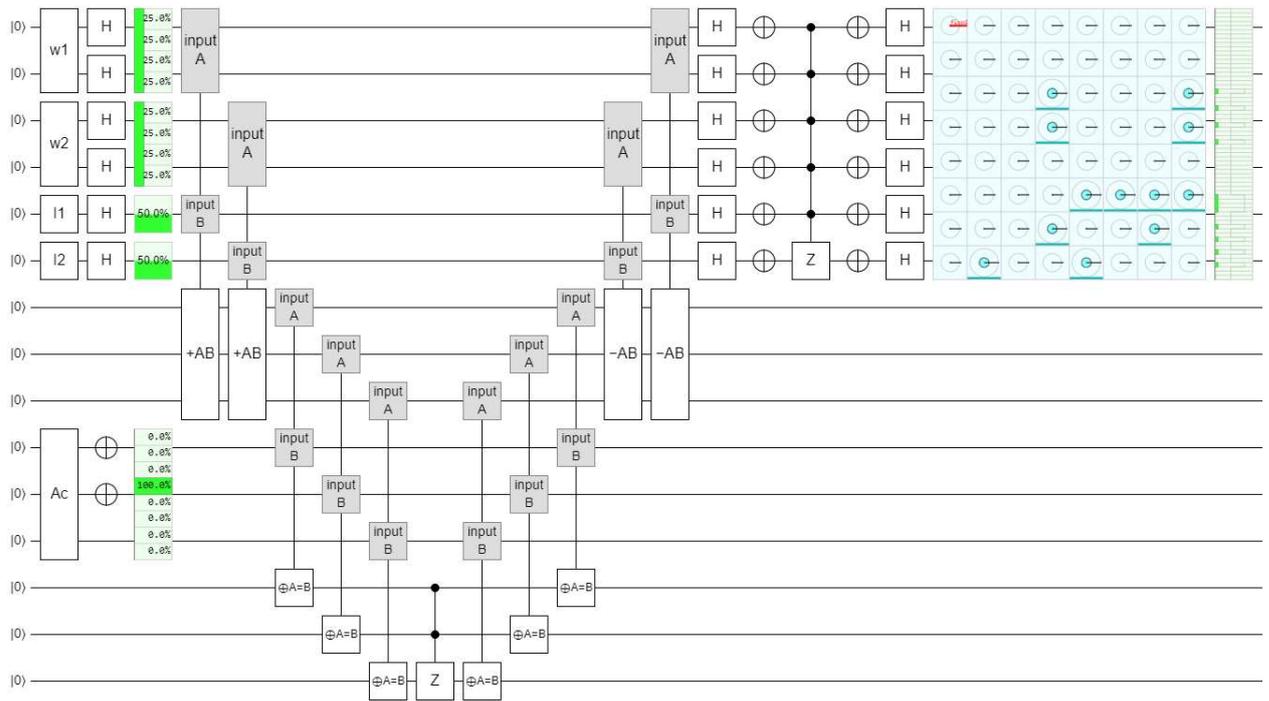

*Fig. 6. Circuit for obtaining all minimal sets of values $I_1$, $I_2$, $w_1$, $w_2$, suitable for training the perceptron.*

The circuit on Fig. 6 allows to obtain all possible combinations of input neuron and weight values that minimally satisfy the activation function: $I_1 w_1 + I_2 w_2 = Ac$. This method can further speed up and simplify the training process, or be used for debugging purposes to get the neural network's response to all possible input values.

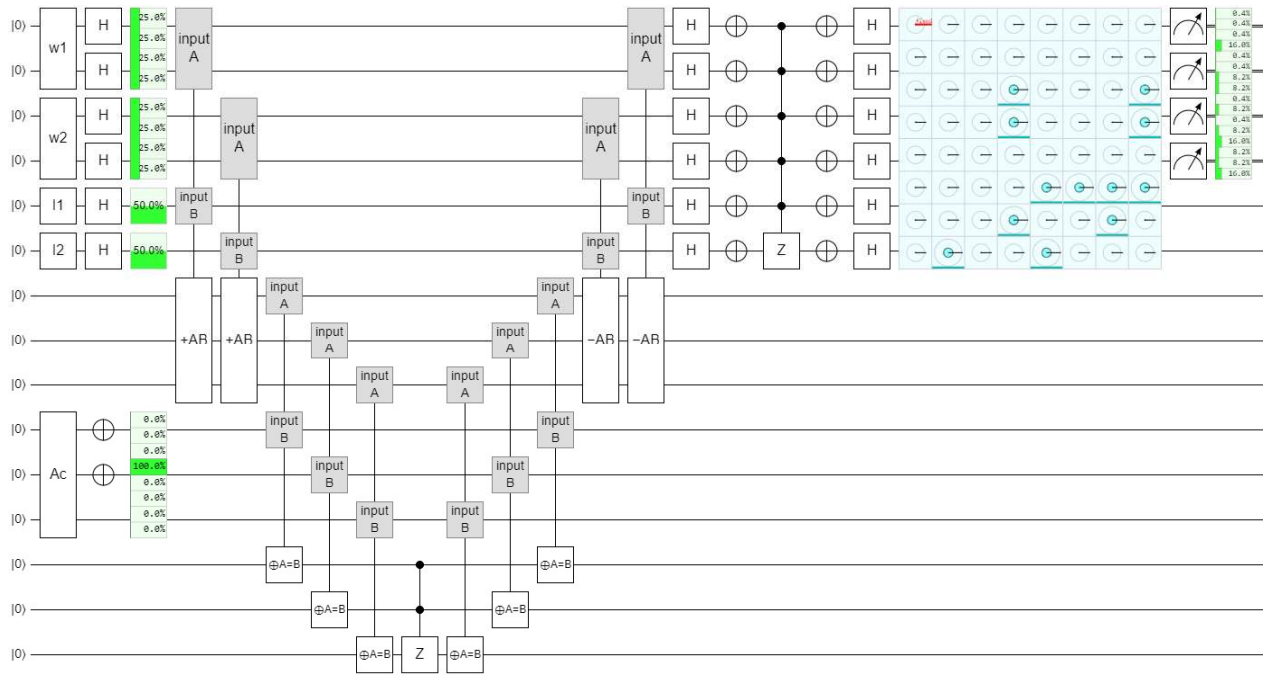

*Fig. 7.a. The order of performing weight measurement for obtaining the distribution of output values $w_1$, $w_2$ in sets of classical bits*

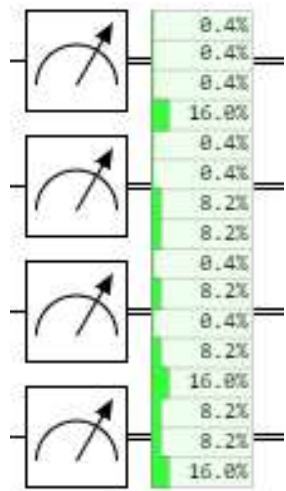

*Fig. 7.b. Distribution of the output values $w_1$, $w_2$ in sets of classical bits, calculated by quantum simulator "Quirk"*

By measuring only the weights of the synapses after each run of the circuit on Fig. 7.a, it is possible to obtain a distribution of the output weight values (Fig. 7.b), from which the weights with the highest probability of being preserved after measurement could be most suitable for training the neural network.

## Conclusion

This paper demonstrates concepts for constructing quantum circuits for training neural networks (Fig. 2, 5) by utilizing the principles of Grover's quantum search algorithm. Based on these concepts, quantum circuits were built for training an example of a simple neural network (Fig. 3, 4, 6). With more available quantum bits and by using the described approach, it is possible to train more complex neural networks.

With further modification of the presented circuits and using more quantum bits, it is possible to create an algorithm that outputs all values that satisfy the inequality $I_1 w_1 + I_2 w_2 \geq Ac$. This can potentially automate this neural network training process.

Further application of Grover's Search Algorithm in machine learning could help significantly speed up and improve the training process of large neural networks [5].

The method for finding solutions to equations or inequalities with many unknown variables by using Grover's algorithm in itself could be useful in mathematical modeling and analyzing responses of complex systems in variable operating conditions by identifying vulnerabilities that are difficult to predict. Also, such calculations could prove useful in improving efficiency and accuracy of decision support systems and the development of artificial intelligence.

**Author details**

**Andrey Vladimirovich Ostroukh**, Russian Federation, full member RAE, Doctor of Technical Sciences, Professor, Department «Automated Control Systems». State Technical University – MADI, 125319, Russian Federation, Moscow, Leningradsky prospekt, 64. Tel.: +7 (499) 151-64-12. http://www.madi.ru , email: ostroukh@mail.ru , ORCID: https://orcid.org/0000-0002-8887-6132

**Cesar Borisovich Pronin**, Russian Federation, PhD student, Department «Automated Control Systems». State Technical University – MADI, 125319, Russian Federation, Moscow, Leningradsky prospekt, 64. Tel.: +7 (499) 151-64-12. http://www.madi.ru , email: caesarpr12@gmail.com , ORCID: https://orcid.org/0000-0002-9994-1032